\documentclass[onecolumn,amsmath,amssymb,12pt,superscriptaddress,nofootinbib]{revtex4}

\usepackage[latin1]{inputenc}
\usepackage[T1]{fontenc}
\usepackage[english]{babel}
\usepackage{amsthm}
\usepackage{amssymb}
\usepackage{amsmath}
\usepackage{amsthm}
\usepackage[]{graphicx}
\usepackage[]{subfigure}
\usepackage{tensor}
\usepackage{color}
\usepackage{cancel}
\usepackage{setspace}
\usepackage{fancyhdr}
\usepackage{framed}
\usepackage{braket}
\usepackage{tikz}
\newcommand{\be}{\begin{equation}}
\newcommand{\ee}{\end{equation}}
\usepackage[bookmarks,linktocpage, colorlinks=true, plainpages = false, citecolor = blue,  linkcolor=blue, urlcolor = blue, filecolor = blue]{hyperref}

\usepackage[title]{appendix}

\begin{document}


\title{Conformal cosmological black holes: Towards restoring determinism to 
Einstein theory \vspace{.3in}}

\author{Fay\c{c}al Hammad} \email{fhammad@ubishops.ca} 
\affiliation{Department of Physics and Astronomy \& STAR Research Cluster, Bishop's University, 2600 College Street, Sherbrooke, QC, J1M~1Z7 Canada} 
\affiliation{Physics Department, Champlain 
College-Lennoxville, 2580 College Street, Sherbrooke,  
QC, J1M~0C8 Canada}

\affiliation{D\'epartement de Physique, Universit\'e de Montr\'eal,\\
2900 Boulevard \'Edouard-Montpetit,
Montr\'eal, QC, H3T 1J4
Canada}

\author{Dilek K. \c{C}iftci}
\email[]{dkazici@nku.edu.tr}
\affiliation{Department of Physics and Astronomy \& STAR Research Cluster, Bishop's University, 2600 College Street, Sherbrooke, QC, J1M~1Z7
Canada} 
\affiliation{Department of Physics, Nam{\i}k Kemal University, 
Tekirda\u{g}, Turkey}

\author{Valerio Faraoni}
\email[]{vfaraoni@ubishops.ca}
\affiliation{Department of Physics and Astronomy \& STAR Research Cluster, Bishop's University, 2600 College Street, Sherbrooke, QC, J1M~1Z7
Canada}

\begin{abstract}
\vspace{.3in} \noindent A widespread solution-generating technique of general relativity consists 
of conformally transforming known ``seed'' solutions. It is shown that 
these new solutions always solve the field equations of a 
pathological Brans-Dicke theory. However, when interpreted as 
effective Einstein equations, those field equations 
exhibit, in the case of a cosmological ``background'', an induced 
imperfect fluid as an additional effective source besides the original 
sources of the ``seed'' solutions. As an application of this feature,  the 
charged 
non-rotating Thakurta black hole, which is conformal 
to Reissner-Nordstr\"om,  is used to demonstrate the 
fragility of the inner Cauchy horizon when this black hole is embedded in 
the universe (even accounting for the separation of black hole and Hubble 
scales). Furthermore, we show that the charged McVittie spacetime, 
although not conformal to any GR solution, represents a 
charged black hole embedded in a cosmological ``background'' with   
varying Hubble parameter that does not exhibit a real Cauchy 
horizon. These 
arguments speak in favor of restoring determinism to Einstein theory, 
which was questioned in recent research.

\end{abstract}
\maketitle

\tableofcontents
\section{Introduction}\label{sec:1}
\setcounter{equation}{0}
The most general spherically symmetric and asymptotically flat solution of the coupled Einstein-Maxwell equations is the Reissner-Nordstr\"om (RN)  geometry describing a charged black hole. Excluding the extremal and super-extremal cases, this static solution has a null event horizon (the outermost black hole horizon) which encloses a null Cauchy horizon. Cauchy horizons are surfaces through which the geometry can be continued but cannot be predicted by prescribing regular initial data. In other words, they are null hypersurfaces that constitute the boundary of the domain of validity of the Cauchy problem for spacetime. The existence of such a boundary, avoided only by the strong cosmic censorship conjecture \cite{Penrose}, implies the loss of determinism in such a spacetime.  Therefore, the initial value problem of vacuum GR fails in the interior of a charged black hole and the theory ceases to be predictive and deterministic, a completely unacceptable shortcoming for any fundamental physical theory or for its solutions. Realistic astrophysical black holes are not charged nor static:  they are electrically neutral and they rotate. However, static charged black holes have been used in the recent Ref.~\cite{Cardosoetal} as toy models for realistic black holes. Fortunately, there is a long history of indications that the Cauchy horizon inside the RN black hole is an artifact of the perfect symmetries of the latter: it is fragile and it disappears when these symmetries are broken or the RN solution is perturbed. Specifically, photons arriving to the Cauchy horizon from larger radii are infinitely blueshifted (a phenomenon known as {\it mass inflation}) and a mass inflation singularity develops when this phenomenon is taken into account \cite{PoissonIsrael90}. The Cauchy horizon is then unstable with respect to perturbations of the RN solution, which decay outside the event horizon but grow in the region inside of it because of the infinite blueshift, transforming the Cauchy horizon into a singularity through which the spacetime cannot be continued. In Ref.~\cite{Cardosoetal} (see also Ref.~\cite{Reall}) it was pointed out, through a clever study of the quasinormal modes, that adding a positive cosmological constant $\Lambda$ to the picture, the resulting Reissner-Nordstr\"om-de Sitter (RNdS) solution of the Einstein equations exhibits a decay rate of the perturbations outside the black hole horizon which is  quite different from that of the RN black hole (exponential instead of power-law \cite{Cardosoetal}). Since the decay rate outside the black hole horizon is tied to mass inflation near the Cauchy horizon, the latter is stabilized by the cosmological constant and determinism is again in jeopardy. This fact is worrysome since, ultimately, no black hole is isolated but it is embedded in the universe and the asymptotics are not Minkowskian, but cosmological. Therefore, the RNdS model is a more realistic model of a black hole than a RN one and the introduction of $\Lambda$ in \cite{Cardosoetal} is well justified. Even though the asymptotics are usually neglected for astrophysical black holes evolving on temporal and spatial scales much smaller than the Hubble radius, this cannot always be done in problems of principle, as Ref.~\cite{Cardosoetal} shows, because even a tiny cosmological constant can make a profound difference. The conclusions of Ref.~\cite{Cardosoetal} have been challenged in Refs.~\cite{Hod1,Hod2,Hod3,Diasetal}. In Ref.~\cite{Hod1,Hod2,Hod3} it is pointed out that scalar field perturbations around a charged black hole necessarily involve a charged scalar field, and that its decay rate outside the black hole horizon is not altered with respect to the RN case. Ref.~\cite{Diasetal} studies instead electrically neutral but rotating black holes in a de Sitter background and shows that the Cauchy horizon is again unstable for this more realistic situation. While the debate continues \cite{next,Dafermos,ReallAug18,MoTianWangZhang}, Ref.~\cite{CardosoAug2018} points out persistent evidence of the loss of determinism in a  finite region of parameter space. Here we propose an independent approach and we point out a different, non-perturbative way in which the Cauchy horizon is destroyed by modifying the black hole  model to make it more realistic. In fact, although the perturbations of the RNdS geometry described by quasinormal modes help break the symmetries, the real universe is not described by an exact de Sitter model. While de Sitter is the late-time attractor of many dark energy models attempting to explain the current acceleration of the universe (including that caused by a cosmological constant) \cite{AmendolaTsujikawa}, and de Sitter space may ultimately turn out to be its final asymptotic state, the real universe contains dark and ordinary matter, radiation, neutrinos and other forms of mass-energy and is not completely empty. The matter stress-energy tensor $T_{ab}$ in the right-hand side of the Einstein equations with cosmological constant $\Lambda$\footnote{Throughout this paper, we use units in which Newton's constant $G$ is unity and we use the notations of Ref.~\cite{Waldbook}.}
\begin{equation}\label{0}
G_{ab} \equiv {\cal R}_{ab}-\frac{1}{2} \, g_{ab} {\cal R} =
8\pi T_{ab}-\Lambda g_{ab} 
\end{equation}
cannot be neglected entirely. As a consequence, the cosmological model 
describing our universe is not a pure de  Sitter space, but rather a FLRW 
one. Then, a better model of a black hole with non-Minkowskian asymptotics is one in which this 
object is embedded in a non-static FLRW universe. Strictly speaking, though, real black holes 
exist in the dynamic backgrounds of surrounding matter (stars, star clusters, galaxies, etc) 
that create much larger curvatures than the cosmological backgrounds. Therefore, the reader 
should keep in mind that while the simple picture of a black hole used here is certainly more 
general and closer to that of realistic black holes in nature than the static ones, this still 
constitutes only a working toy model.

There are immediately two 
problems arising 
with such simplified models, though. First, while the RNdS geometry is the 
unique spherical, static, and asymptotically de Sitter solution of the 
coupled  Einstein-Maxwell equations, there is no unique solution with 
dynamical FLRW asymptotics. While a few exact solutions of the Einstein 
equations (and their charged versions) are known, they are  special 
and 
they usually suffer from some physical pathology (see Ref.~\cite{mybook} 
for 
a review). We argue that 
they are still better models of charged black holes than the RNdS space in 
the sense that they model the cosmological asymptotics in a  general 
(instead of locally static) way. 

The second 
problem is that, while in stationary situations (such as for the 
RNdS model) black hole horizons are static and null surfaces, for  
dynamical black holes one must consider instead {\em apparent} horizons 
(AHs), 
which are dynamical and are spacelike or timelike. However, AHs are 
foliation-dependent, as exemplified 
dramatically by the fact that in the Schwarzschild spacetime  there exist 
foliations without AHs \cite{WaldIyer1,WaldIyer2}. This 
problem is somehow 
alleviated by the recent realization that, in spherical symmetry, 
all spherically symmetric foliations (to 
which we restrict here) possess the 
same AHs \cite{Ellis}. In any case, the recent detections of 
gravitational waves from black hole mergers by the {\em LIGO} 
interferometers \cite{LIGO1,LIGO2,LIGO3,LIGO4} are based in  
an essential way on the use of marginally trapped surfaces and AHs. In 
fact, due to the low signal to noise ratio,  
gravitational wave signals are matched to banks of templates for the 
gravitational waveforms, which are produced by numerical simulations 
identifying black holes with their apparent, not event, horizons. Event 
horizons are essentially useless for this practical task. The new and 
promising gravitational wave science is based on AHs  when 
these waves are generated by mergers of black holes with other objects.

Keeping in mind the two {\em caveats} above, one can nevertheless study 
particular solutions of GR describing charged black holes embedded in 
FLRW universes as more general toy models than RNdS.  We discuss two  
examples in which, changing the background from static Minkowski or de 
Sitter to FLRW,  
the inner Cauchy horizon disappears. This is a further indication 
that the Cauchy horizon is very fragile and is not expected to occur in 
nature, helping restore 
determinism to GR.

Another topic in gravitational physics, that might seem  
outside the purpose of the present paper,  is conformal transformations.  
Conformal transformations of the spacetime metric play an important role 
in general relativity (GR) in the study of conformal infinity and in  
the 
construction of Penrose-Carter diagrams \cite{Waldbook}, in 
alternative theories of gravity where different conformal frames (the 
Jordan and the Einstein frames) provide alternative representations of 
these theories \cite{Dicke,FujiiMaeda,mybook}, in highlighting the 
true nature of some of the quasi-local definitions of mass in  GR and in  
scalar-tensor gravity \cite{Faycal1,Faycal2}, in examining from a different angle the thermodynamics of black holes \cite{FEP1} and the thermodynamics of classical spacetime in general \cite{FD}, and in investigating the relation between the energy conditions and wormholes \cite{FEP2}. More important for us, however, is the fact that conformal transformations are also used in GR as a technique to generate new analytic  solutions of the Einstein equations starting from known ones, particularly in the case of electrovacuum \cite{Szekeres,Thakurta,vandenBergh1,vandenBergh2,vandenBergh3,vandenBergh4,vandenBergh5,vandenBergh6,HMN, Fonarev,SultanaDyer,McClureDyer2006,RodriguesZanchin,MelloMacielZanchin}, 
but also in the presence of fluids  \cite{LorangerLake,Moradpouretal2015,Hansraj1,Hansraj2,Hansraj3}. This technique allows, in fact, to change the background of a 
given GR solution from the static Minkowski or de Sitter to the needed FLRW.
 This 
technique has even more potential in the context of scalar-tensor gravity as it allows to obtain the general vacuum and electrovacuum static, spherically symmetric solutions, including the Brans-Dicke theory as a special case \cite{Bronnikov}. This technique has subsequently been used \cite{Bekenstein,CliftonMotaBarrow1,CliftonMotaBarrow2,NMCreferencesEllis1,NMCreferencesEllis2,NMCreferencesEllis3,Shawnthesis,BDsymmetry} 
to generate the general spherically symmetric 
static solution of vacuum Brans-Dicke theory from a known 
general solution of GR 
(see, {\em e.g.}, \cite{VFAT} and the references therein). For our
present purpose of tackling the issue of determinism loss
in Einstein's theory, however, we restrict ourselves to GR.

Let the spacetime metric $g_{ab}$ be a solution of the Einstein 
equations~(\ref{0}) without the cosmological constant. Then, it is {\em a 
priori} possible that the conformally related metric 
\be \label{1}
\tilde{g}_{ab}=\Omega^2 \, g_{ab} \,, 
\ee
where $\Omega$ is a nowhere-vanishing, regular, conformal factor is still 
a solution of the Einstein equations. In 
order for this new solution to 
be of any physical interest, however,
 the conformal factor $\Omega$ must be chosen 
judiciously.   Over the years, there has been increasing 
interest in generating solutions of the Einstein equations which describe 
black holes embedded in cosmological ``backgrounds''.\footnote{We use 
quotation marks because, due to the non-linearity of the field 
equations, one cannot split a metric into a ``background'' 
and a ```deviation'' from it in a  covariant way (apart from algebraically special geometries, such as the Kerr-Schild ones).} In these cases, the 
metric $g_{ab} $ usually describes a black hole (Schwarschild, Kerr, or 
their 
charged generalizations) and 
the conformal factor $\Omega$ is chosen as the scale factor of a 
spatially flat Friedmann-Lema\^itre-Robertson-Walker (FLRW) universe, 
which 
comes to constitute the cosmological ``background''. This procedure has 
generated the Thakurta \cite{Thakurta}, Sultana-Dyer \cite{SultanaDyer}, 
McClure-Dyer \cite{McClureDyer2006}, and other solutions such as  
spherical perfect fluid solutions \cite{LorangerLake,Hansraj1,Hansraj2,Hansraj3, Moradpouretal2015}. 

The new metric $\tilde{g}_{ab}$ is not  
a solution of the Einstein equations with the same form of matter source for 
which the original metric 
$g_{ab}$ is a solution, though. In fact, under the conformal 
transformation~(\ref{1}), the Ricci tensor changes according 
to \cite{Synge}
\begin{equation} 
\tilde{{\cal R}}_{ab} = {\cal R}_{ab} + 
 \frac{4\nabla_a\Omega\nabla_b\Omega}{\Omega^2}-
\frac{2\nabla_a\nabla_b\Omega}{\Omega}
-g_{ab}
\left[\frac{\Box\Omega}{\Omega}+\frac{\nabla_e\Omega\nabla^e\Omega}{ 
\Omega^2}\right],
\label{2}
\end{equation}
while the trace of this equation gives
\be
\tilde{ {\cal R}} = \Omega^{-2} \left( {\cal R}  
-\frac{6\Box\Omega}{\Omega}\right) 
\,,\label{2bis}
\ee
so that Eq.~(\ref{0}) becomes
\begin{align}
\tilde{G}_{ab} &= \kappa T_{ab} -\frac{2}{\Omega} \left( \nabla_a 
\nabla_b \Omega -g_{ab} \Box \Omega \right)+ \frac{1}{\Omega^2} \left( 
4\nabla_a \Omega \nabla_b \Omega-g_{ab} \nabla_c \Omega \nabla^c \Omega 
\right)\nonumber\\
&\equiv \kappa \left(T_{ab}+ T_{ab}^{(\Omega)}\right).  
\label{modifiedEFE}
\end{align}
Thus,  a vacuum solution $g_{ab}$ (with $ {\cal R}_{ab}=0$) is transformed 
into a non-vacuum one with $\tilde{ {\cal R}}_{ab}\neq 0$. The derivatives 
of the scale 
factor $\Omega$ act as an effective form of matter in the right-hand side 
of the Einstein 
equations. Since generating new solutions in this way amounts to the 
``Synge procedure'' of imposing the form of the metric 
and then running the Einstein equations to determine the matter that makes 
the chosen metric a solution, there is {\em a priori} little hope that 
this artificially created effective  matter $T_{ab}^{(\Omega)}$ will be 
physically meaningful. Indeed, the right-hand side of the tilded Einstein  
equations~(\ref{modifiedEFE}) 
contains, in addition to 
``standard'' terms quadratic in the gradient $\nabla_a \Omega$, terms 
linear in the second derivatives $\nabla_a \nabla_b \Omega$ and $\Box 
\Omega$. These terms make the sign of the effective energy density 
undefined and $T_{ab}^{(\Omega)}$ will not, in general, satisfy any 
energy condition.  In fact, the ``cosmological black hole'' geometries 
generated this way are 
often reported to exhibit negative energy densities in certain spacetime 
regions \cite{SultanaDyer,McClureDyer2006}.  Furthermore, the new solutions thus obtained might actually exhibit singularities that may or may not be present in the original solution. Indeed, as can be seen from (\ref{2bis}), 
the new Ricci scalar $\tilde{{\cal R}}$ might possess a singularity whenever the inverse metric component in $\Box\Omega=g^{ab}\nabla_a\nabla_b\Omega$ 
becomes singular. 

Nevertheless, these solutions of GR are still seen as 
interesting at least as toy models of black holes, and they are put to 
use, for example,  
in recent investigations of the Hawking radiation and thermodynamics of 
dynamical black holes (\cite{SaidaHaradaMaeda, myHawking,Majhi1,Majhi2,Majhi3,Majhi4}, see 
also the related references \cite{SasakiDeruelle,FaraoniNielsenCQG,NielsenFirouzjaee}). Furthermore,  the Husain-Martinez-Nu\~nez solution of 
GR 
\cite{HMN} is 
conformal to the Fisher solution (with the scale 
factor of the FLRW ``background'' universe as the conformal 
factor), but has 
as the matter source 
a canonical, minimally coupled, free scalar field which satisfies both
 the weak 
and null energy conditions. The same can be said about its generalization 
in which the scalar acquires an exponential potential, known as the 
Fonarev solution \cite{Fonarev}. It may even happen that an 
unphysical solution of the Einstein equations is conformally 
transformed into a  physically 
interesting one, as is the case for the fluid spheres of Ref.~ 
\cite{LorangerLake}.  On the other hand, whenever $g_{ab}$ is an 
(electro)vacuum 
solution of the Einstein equations, the geometry 
$\tilde{g}_{ab}=\Omega^2 \, g_{ab}$ can always be seen as 
a solution of a Brans-Dicke theory with Brans-Dicke coupling 
$\omega=-3/2$. This theory is known to be pathological,\footnote{See, however, Refs.~\cite{pathological1,pathological2,pathological3,pathological4,pathological5,pathological6,pathological7} exploring it.} in the sense that the Brans-Dicke field $\phi=\Omega^{-2}$ is non-dynamical and, more important for us, the Cauchy problem  becomes ill-posed. The pathology is then just a reflection of the fact  that the conformal factor $\Omega$ is forced 
arbitrarily into the geometry by imposing that $\tilde{g}_{ab}$ describe a 
central object in a cosmological ``background'', and this fact is obviously bound to have consequences. 

The outline of the rest of this paper is as follows. 
Section~\ref{sec:2} discusses these general aspects of obtaining new 
solutions of the Einstein equations via conformal transformations;  
Sec.~\ref{sec:3} discusses an ambiguity present in the literature about 
the non-rotating Thakurta and Sultana-Dyer cosmological black 
hole geometries, which are of the kind described above. The 
following section focuses on the non-rotating Thakurta solution, 
which is of special interest in both GR and scalar-tensor 
gravity. We use this conformally created solution there 
as a counterexample to study the (absence of the) inner 
Cauchy
horizon in cosmological black holes. Section~\ref{sec:5} 
proposes yet another cosmological black hole which cannot be built 
by a conformal transformation, as a 
useful example in the debate about inner Cauchy horizons in GR black 
holes.  Finally, Sec.~\ref{sec:6} contains the conclusions. 

\section{GR seed geometries and $\omega=-3/2$ Brans-Dicke gravity} \label{sec:2} 
\setcounter{equation}{0}

In this section we expose the physical features we gain by using 
conformal transformations to construct a new solution from a given GR 
solution. In addition, we also describe the status of such a process 
within the framework of Brans-Dicke theory.

Assume that $g_{ab}$ is an (electro)vacuum solution of the Einstein 
equations~(\ref{0}) obtained from the Einstein-Hilbert action
\be
S = \int d^4 x \sqrt{-g} \left[ \frac{1}{2\kappa } \left( {\cal R} 
-\Lambda 
\right) 
+ {\cal L}_{(m)} \left[ g_{ab}, \psi \right]  \right] \,, 
\label{EH}
\ee
where $\Lambda$ is the cosmological constant and ${\cal L}_{(m)}\left[ 
g_{ab}, \psi \right] $ is the matter Lagrangian, with $\psi$ denoting 
collectively the matter fields.  Consider the conformal metric 
$\tilde{g}_{ab}=\Omega^2 \, g_{ab}$: by using Eqs.~(\ref{2}), 
(\ref{2bis}), and 
\be
\sqrt{-\tilde{g}}=\Omega^4 \, \sqrt{-g}\,,\qquad
\frac{ \Box \Omega}{ \Omega^3} = 
\frac{ \tilde{\Box} \Omega}{ \Omega} -\frac{2\tilde{g}^{ab} 
\tilde{\nabla}_a \Omega \tilde{\nabla}_b \Omega}{ \Omega^2} \,,
\ee
and introducing the scalar field $\phi =\Omega^{-2}$ 
(which will become a Brans-Dicke scalar), one easily obtains
\begin{equation}
\sqrt{-g} \, {\cal R} = \sqrt{-\tilde{g}} \left( \phi \tilde{ {\cal R}} 
+\frac{3}{2\phi} \, \tilde{g}^{ab} \tilde{\nabla}_a\phi 
\tilde{\nabla}_b\phi \right)
-3\partial_c 
\left( \sqrt{-\tilde{g}} \, \tilde{g}^{ac} \partial_{a} \phi \right) 
\,. \label{obtained}
\end{equation}
In terms of $\tilde{g}_{ab}$ and $\phi$, the Einstein-Hilbert 
action~(\ref{EH}) becomes
\begin{equation}
S = \int d^4 x \sqrt{-\tilde{g} } \Bigg\{ \frac{1}{2\kappa} \left[ 
\phi 
\tilde{ {\cal R}} 
+\frac{3}{2\phi} \, \tilde{g}^{ab} \tilde{\nabla}_a \phi \tilde{\nabla}_b 
\phi -V(\phi) \right]
+ \tilde{ {\cal L}}_{(m)} \left[ 
\tilde{g}_{ab}, \psi \right] \Bigg\}  \label{BDaction}
\,,
\end{equation}
where the total divergence in the right-hand side of 
Eq.~(\ref{obtained}) 
(which only contributes a boundary term to the 
action), has been dropped, 
 the cosmological constant has become the mass potential
\be
V(\phi) = \frac{\Lambda}{2\kappa} \, \phi^2  \equiv \frac{  
\mu^2 \phi^2}{2} \,,
\ee
and the matter Lagrangian density reads now
$\sqrt{-\tilde{g}}\tilde{ {\cal L}}_{(m)}=\phi^{-2} \sqrt{-g}{\cal L}_{(m)} \left[ \phi \tilde{g}_{ab}, \psi \right]$. 
This is a (Jordan frame) Brans-
Dicke action \cite{BransDicke}  with 
coupling parameter $\omega=-3/2$. Its variation with respect to 
$\tilde{g}^{ab}$ and $\phi$ generates the field equations
\begin{equation}
\tilde{ {\cal R} }_{ab} - \frac{1}{2}\, \tilde{g}_{ab}  \tilde{ {\cal R} }   
=   \frac{\kappa \tilde{T}_{ab} }{\phi}  -\frac{3}{2\phi^2} \left( 
\tilde{\nabla}_a \phi 
\tilde{\nabla}_b \phi -\frac{1}{2} \, \tilde{g}_{ab} 
\tilde{\nabla}_c \phi \tilde{\nabla}^c \phi \right) +\frac{1}{\phi} \left( \tilde{\nabla}_a\tilde{\nabla}_b \phi 
-\tilde{g}_{ab} \tilde{\Box}\phi \right) 
-\frac{V}{2\phi}\, 
\tilde{g}_{ab},\label{2-7}
\end{equation}
\be
\tilde{\Box}\phi = \frac{\phi}{3} \left( \tilde{ {\cal R}} 
-\frac{dV}{d\phi} 
\right) +\frac{1}{2\phi} \tilde{g}^{cd}
\tilde{\nabla}_c \phi \tilde{\nabla}_d \phi \,.\label{2-8}
\ee
The tilded energy-momentum tensor $\tilde{T}_{ab}$ in (\ref{2-7}) is 
related to the original energy-momentum tensor $T_{ab}$ by 
$\tilde{T}_{ab}=\phi T_{ab}$. In other words, what appears on the right-hand side of the equations is $\kappa T_{ab}$, {\it i.e.}, the original energy-momentum tensor of the sources. 

The field equations~(\ref{2-7}) can be interpreted as usual 
Einstein field equations, but with an additional effective-matter source.  
In this context, notice the important fact, to which we 
shall come back below, that the absence of a radial matter flow in the 
original spacetime, $T_{01}=0$, does not prevent such a radial flow from 
emerging in the new frame even with a conformal 
factor which depends only on time. This general pattern stems from 
the fact that the $(0,1)$ component  
of the second derivative $\tilde{\nabla}_{a}\tilde{\nabla}_{b}\phi$ is 
not zero. Physically, this could be understood as the result of the 
original radial dependence of the metric transformed into an 
effective radial flow due to the stretching of spacetime in a 
time-dependent way. This is illustrated by the expression 
(\ref{uAndq2}) of the energy flux  density $q_a$, which would  
identically vanish only for a time-independent conformal factor. 

In general, just as there is an induced energy flow in the form of a 
non-vanishing $T_{01}^{(\Omega)}$, the field equations for the matter 
fields $\psi$ also acquire extra terms due to the new form 
 of the matter Lagrangian, which picks up an explicit 
dependence on the scalar field $\phi$. This fact does not, however, arise 
for conformally invariant matter fields, as is the case for  
the Maxwell field whose 
Lagrangian density $- \sqrt{-g} \, F_{ab}F^{ab}/4 $ is invariant under  
conformal transformations. As such, the Maxwell equations in vacuo are 
also conformally invariant. We shall come back to this 
observation in  
Sec.~\ref{sec:4}, where we examine the charged Thakurta black hole. 

Now, Brans-Dicke theory with the particular 
value $-3/2$ of the $\omega$-parameter   
is known to be pathological: the Brans-Dicke scalar $\phi$ (corresponding 
approximately to the inverse of the gravitational coupling) is not 
dynamical. In fact, by taking the trace of Eq.~(\ref{2-7}),
\be
\tilde{ {\cal R}} = -\frac{\kappa \tilde{T}}{\phi} -\frac{3}{2\phi^2} \, 
\tilde{g}^{ab} 
\tilde{\nabla}_a \phi \tilde{\nabla}_b \phi 
+\frac{3\tilde{\Box}\phi}{\phi} 
+\frac{2V}{\phi} \,,
\ee
and substituting it into Eq.~(\ref{2-8}) reduces the latter to $ 
\tilde{T}=0 $, and therefore $T=0$, 
which is identically satisfied with a conformally invariant 
form of matter in the original frame and, in particular, (electro)vacuum. 
The usual  Brans-Dicke dynamical equation for 
$\phi$ is thus completely lost and this field is not even subject to a 
first 
order constraint, becoming completely arbitrary. 
Correspondingly, the Cauchy problem for $\omega=-3/2$ is ill-posed 
(\cite{Salgado}, see also 
\cite{Noakes,others1,others2,Tremblay1,Tremblay2,Tremblay3,reviews1,reviews2,reviews3}). This feature resurfaced 
recently in the literature 
with the revival of 
Palatini $f( {\cal R})$ gravity as an alternative to dark energy to 
explain the 
current acceleration of the universe, because this theory is 
equivalent to $\omega=-3/2$ Brans-Dicke gravity with a special potential 
\cite{reviews1,reviews2,reviews3}. 

These properties are not surprising because, while the 
geometry $g_{ab}$ solves the Einstein equations (with no matter or with  
just the Maxwell field), the conformal factor $\Omega$ is completely 
arbitrary and is introduced {\em ad hoc} without being required 
to satisfy 
any rule or physical equation. Requiring $\Omega$ to coincide with the 
scale factor of a ``background'' FLRW universe does introduce some 
physics into this picture, but this is still an artificial way to 
force a geometry to do what we want. While the goal of the 
transformation~(\ref{1}) is to generate new solutions $\tilde{g}_{ab}$ 
{\em of GR} with some desired properties, these can always be seen also as 
solutions of the (pathological) $\omega=-3/2$ Brans-Dicke gravity, 
possibly with a mass potential.

\section{Relations between conformal GR solutions}
\label{sec:3}
\setcounter{equation}{0}
Being concerned here specifically with the non-rotating Thakurta cosmological 
black hole, we discuss in this section an ambiguity 
present in the literature about such a black hole and the Sultana-Dyer 
one, which is also obtained by conformally transforming 
the same GR seed geometry.

Let the metric $g_{ab}$  be an (electro)vacuum solution of the Einstein 
equations which can be expressed in various coordinate systems. Let 
$g_{\mu\nu}$ and $g_{\mu'\nu'}$  denote, respectively, the metric 
components in two coordinate systems (for example, consider the 
Schwarzschild metric in Schwarzschild, isotropic, Kerr-Schild, or  
Eddington-Finkelstein coordinates \cite{Poisson}). By conformally 
transforming $g_{ab}$ with a conformal factor $\Omega$, one obtains the 
two 
expressions $\tilde{g}_{\mu\nu}=\Omega^2 g_{\mu\nu}$ and  
$\tilde{g}_{\mu'\nu'}=\Omega^2 g_{\mu'\nu'}$ of the same metric.  This 
stems from the one-to-one character 
of the  conformal transformations (\ref{1}) in the case of 
(electro)vacuum, as shown in Ref.~\cite{VFAT}. There are, however,  
incorrect claims to the contrary in the literature.  For example, in 
Ref.~\cite{McClureDyer2006}, the Schwarzschild  metric in Schwarzschild 
coordinates $(t,r,\theta,\phi)$ and in 
isotropic coordinates $(t,\rho,\theta,\phi)$, respectively,  
\begin{eqnarray}
ds^2_{(S)} &=&  -\left( 1-\frac{2m}{r} \right) dt^2 + \frac{dr^2}{ 
1-\frac{2 m}{r}} +r^2 d\Omega_{(2)}^2 \label{S1}\\ 
&& \nonumber\\
&=& -\left( \frac{1-\frac{m}{2\rho}}{1+\frac{m}{2\rho} } \right)^2 
dt^2 + 
\left( 1 +\frac{m}{2\rho} \right)^4 \left( d\rho^2 + \rho^2 
d\Omega_{(2)}^2  \right)
\label{S2}
\end{eqnarray}
(where $d\Omega_{(2)}^2 = d\theta^2 +\sin^2 \theta \, d\varphi^2$ is the 
line element on the unit 2-sphere) is conformally transformed. If the line 
element in the 
form~(\ref{S1}) is  used, one obtains the non-rotating Thakurta metric
\be
ds^2_{(T)}= a^2(t) \left[ -\left( 1-\frac{2m}{r} \right) dt^2 +
\frac{dr^2}{ 1-\frac{2m}{r}} +r^2 d\Omega_{(2)}^2 \right] 
\label{NRT-Schwarzschild}
\ee
where $a(t) $ is the scale factor of the FLRW ``background'' universe 
into which the Schwarzschild black hole gets embedded. In 
Ref.~\cite{McClureDyer2006}, the line element 
\begin{equation}
ds^2 =  a^2(t)\Bigg[ 
-\left( \frac{1-\frac{m}{2\rho}}{1+\frac{m}{2\rho} } \right)^2 dt^2
+\left( 1+\frac{m}{2\rho} \right)^4\left( d\rho^2 + \rho^2
d\Omega_{(2)}^2
\right)\Bigg]    \label{NRT-isotropic}
\end{equation}
obtained by conformally transforming the Schwarzschild line element in its 
form~(\ref{S2}) with the same conformal factor $a$,  
is presented as a new GR solution alternative to the Thakurta one. 
However, the usual coordinate transformation 
\be
\rho \rightarrow r=\rho \left( 1+\frac{m}{2\rho} \right)^2  
\ee
turns the line element~(\ref{NRT-isotropic}) 
into~(\ref{NRT-Schwarzschild}).

In contrast with the two forms~(\ref{NRT-Schwarzschild}) and 
(\ref{NRT-isotropic}), the non-rotating Thakurta  and the 
Sultana-Dyer \cite{SultanaDyer} solutions  are genuinely different 
from each other, in spite 
of being both conformal to Schwarzschild because they are generated using 
two different conformal factors in Eq.~(\ref{1}). This fact (remarked in 
Ref.~\cite{MelloMacielZanchin}) is not obvious in the 
coordinate systems normally used in the literature. The Sultana-Dyer line 
element is \cite{SultanaDyer}
\begin{eqnarray}
ds^2_{(SD)} &=& a^2(\tau) \left[ -d\tau^2 +\frac{2 m}{r}(d\tau+dr)^2 
+dr^2 
+r^2 d\Omega_{(2)}^2 
\right] \label{105}\nonumber\\
&&\nonumber\\
& = &  a^2(\tau) \Bigg[ -\left( 1-\frac{2m}{r}\right) d\tau^2 + 
\frac{4 m}{r} \, d\tau dr+ \left( 
1 + \frac{2m}{r}\right) dr^2+r^2 
d\Omega_{(2)}^2 \Bigg]\label{106}
\end{eqnarray}
where $a(\tau)=\tau^2$ and $m>0 $ is  the mass of the original 
Schwarzschild black hole  
\cite{SultanaDyer}. It is already clear from this last expression of the 
Sultana-Dyer metric that the latter is fundamentally different from 
the conformal  Schwarzschild metric~(\ref{NRT-Schwarzschild}) that has a 
conformal factor depending only on time.  To make this difference more 
apparent, let us introduce a new time 
coordinate $t $ defined by 
\be
\tau(t, r ) = t+2m \ln \left| \frac{r}{2m} -1 \right| \,,
\ee
which will be interpreted as the conformal time of the FLRW ``background'' 
universe. Differentiation gives 
\begin{equation}
d \tau = d t + \frac{2 mdr}{r\left( 1- 2m/r \right)}   
\,, \label{trouble}
\end{equation}
and substitution into Eq.~(\ref{106}) turns this line element 
into the diagonal form
\begin{eqnarray}
ds^2 =  a^2(t, r) 
\left[ -\left( 1-\frac{2 m}{r}\right) dt^2 + 
\frac{dr^2}{1-\frac{2 m}{r}}+r^2
d\Omega_{(2)}^2 \right] \,. 
\end{eqnarray}
In these coordinates, the Sultana-Dyer line element is explicitly 
conformal 
to Schwarzschild, with conformal factor 
\be 
\Omega= a(t,r)= \tau^2(t, r)= \left( t + 2m \ln \left| 
\frac{r}{2 m}-1 \right| \right)^2 \,,
\ee
which is clearly different from the conformal factor of the non-rotating 
Thakurta metric~(\ref{NRT-Schwarzschild}), which depends only on time. 
It is specifically this latter metric that is going to serve our 
purpose here. In the next section, we are first going to expose the 
construction and the physical features of the non-charged version in order 
to fully understand the more useful charged one, presented in 
Sec.~\ref{sec:ChargedNRT} and then used to examine the problem of 
the Cauchy horizon in Sec.~\ref{sec:CauchyHorizonNRT}.

\section{Thakurta geometry and strong cosmic censorship}
\label{sec:4}
\setcounter{equation}{0}

\subsection{Uncharged non-rotating Thakurta metric} 

The non-rotating Thakurta geometry has been the subject of 
recent attention \cite{McClureDyer2006,SaidaHaradaMaeda,myHawking,Culetu,Majhi1,Majhi2,Majhi3,Majhi4,MelloMacielZanchin} and 
is related to other exact solutions, hence it deserves a better 
look. As we shall see below, the other expression of the Thakurta line 
element~(\ref{NRT-isotropic}) looks 
superficially  like that of 
the McVittie metric \cite{McVittie}:
\begin{equation}
ds^2=-\left( \frac{1-\frac{m}{2ar}  }{1+\frac{m}{2ar}}\right)^2\,  dt^2+a^2(t)\left( 
1+\frac{m}{2ar} \right)^4\left(dr^2+r^2 d\Omega_{(2)}^2 \right).
\label{McVlineelement}
\end{equation}
\noindent But, while the mass coefficient in the McVittie metric is $ M=m/a(t)$ (with constant $m$), that of the 
line element~(\ref{NRT-isotropic}) is strictly constant. The difference is 
crucial because, allowing $ M$ to be different from its McVittie form 
$ m/a(t)$, implies that there is a (purely spatial) radial energy flux 
with density $q^a$ described by an imperfect fluid term in the matter 
stress-energy tensor \cite{AudreyPRD} 
\be\label{EMTensor}
T_{ab}^{(fluid)}=\left( P+\rho \right) u_a u_b +P g_{ab}+ q_a u_b + q_b 
u_a \,.
\ee
Therefore, instead of a McVittie metric, the non-rotating Thakurta 
solution is  a {\em generalized McVittie}  
geometry of the class presented in Ref.~\cite{AudreyPRD} and studied in 
Ref.~\cite{extraMcVittie1,extraMcVittie2,extraMcVittie2bis,extraMcVittie3}.  The McVittie form $ M=m/a(t)$ of the 
mass  
parameter corresponds to 
the condition ${ G^0}_1=0 $ and, because of the Einstein equations, to 
vanishing radial energy flux ${T^0}_1$ (this is known as the ``McVittie 
condition''). By relaxing the McVittie condition, a radial flux 
$ {T^0}_1$ associated with  an imperfect fluid appears. As remarked 
below Eq.~(\ref{2-8}), we can now understand the origin of this emergent 
radial flow in the non-rotating Thakurta spacetime  as being due to the 
non-vanishing second derivative $\tilde{\nabla}_{0} 
\tilde{\nabla}_{1}a(t)$ of the time-dependent conformal factor $a(t)$. 

Contrary to 
other classes of solutions introduced to describe central objects 
embedded in cosmological spacetimes, in universes  that expand forever or 
in phantom 
cosmologies that end in a Big Rip at a finite future, the generalized 
McVittie class has a late-time attractor  
\cite{McVittieattractor}, which is precisely the non-rotating Thakurta 
solution.\footnote{In Ref.~\cite{McVittieattractor}, the late-time 
attractor was not recognized as a non-rotating Thakurta solution and was 
called ``comoving mass'' solution instead. Similarly,  
Ref.~\cite{CliftonMotaBarrow1,CliftonMotaBarrow2} does not identify the $\omega\rightarrow 
\infty$ limit of its class of Brans-Dicke spacetimes as the Thakurta 
solution.} This geometry is also the limit to GR of a family of  
solutions 
of Brans-Dicke theory found in Ref.~\cite{CliftonMotaBarrow1,CliftonMotaBarrow2}. Furthermore, 
it 
is also a solution of cuscuton gravity (a special 
Ho\v{r}ava-Lifschitz theory \cite{Afshordi1,Afshordi2}) 
and of shape dynamics \cite{genMcVittieHorndeski}.

As follows from the discussion of Sec.~\ref{sec:1}, being 
conformal to the Schwarzschild black hole, the Thakurta metric 
(\ref{NRT-Schwarzschild})
represents a spacetime filled with an artificially created 
effective matter that might exhibit negative energy 
densities in certain spacetime regions. In addition, given that the 
inverse metric $g^{ab}$  of the original Schwarzschild line element used 
to create such a spacetime is singular at the black hole horizon $r=2m$, 
we expect that the non-rotating Thakurta  
spacetime~(\ref{NRT-Schwarzschild})  
 will 
also possess a singularity at that same coordinate location. In fact, 
the coordinate singularity becomes, as we shall see, a true spacetime 
singularity in the conformally mapped geometry. 

Using the action~(\ref{EH}) and defining an  
energy-momentum tensor for the matter part by Eq.~(\ref{EMTensor}), we 
find that to satisfy the Einstein equations~(\ref{0}), the 
components of 
the four-velocity vector and the energy flux density should 
be, respectively,
\begin{eqnarray}
u_a &=& \left(  - a\sqrt{1-\frac{2m }{r}},0,0,0\right)\,,\\
&&\nonumber\\ 
q_a &=&\left( 0,\frac{-m\dot{a}}{4\pi 
a^2r^2\left(1-\frac{2m}{r}\right)^{3/2}},0,0 \right) \, .
\end{eqnarray}
On the other hand, the required energy density and 
pressure of the artificial fluid are found to be, respectively,
\begin{eqnarray}
\rho &=& \frac{3\dot{a}^2}{8\pi a^4(1-2m/r)} \,,\label{RhoP1} \\
&&\nonumber\\
P &=&  \frac{\dot{a}^2-2a\ddot{a}}{
8\pi a^4\left(1- 2m/r \right)}  \,. \label{RhoP2}
\end{eqnarray}
The Ricci scalar of the  Thakurta 
metric~(\ref{NRT-Schwarzschild}), computed directly from 
Eq.~(\ref{2bis}), is 
\be
\mathcal{R} = \frac{  6 \left( \dot{H}+H^2 \right)}{a^2(1-2m/r)}\,.
\ee
As expected, the Ricci scalar is singular  at 
$r=2m$. This singularity translates into a singular 
fluid as both the energy density and pressure 
(\ref{RhoP1}) and (\ref{RhoP2}) diverge there as well.

All the previous results concerning the possibility of a negative
energy density and a singular character of the fluid 
necessary for the creation of the non-rotating Thakurta 
spacetime remain valid in the case of a charged and/or rotating Thakurta 
spacetime.  In what follows, we examine this case in detail.

\subsection{Charged non-rotating Thakurta metric}\label{sec:ChargedNRT}

The charged non-rotating Thakurta (CNRT) metric of a black hole of constant mass $m$ and 
constant charge $Q$ embedded in a cosmological background of scale factor 
$a(t)$ reads
\be\label{ChargedThakurta}
ds_{_{(CNRT)}}^2=a^2(t) \left[-f(r)dt^2 +\frac{dr^2}{f(r)}+r^2 
d\Omega_{(2)}^2 \right] \,,
\ee
where 
\be
f(r)=1-\frac{2m}{r}+\frac{Q^2}{r^2} \,.
\ee
This metric being conformal to the RN metric, and given the conformal 
invariance of the electromagnetic field $F_{ab}$, the corresponding 
expression of the latter for this geometry is  the same as 
the one of the RN spacetime  
$F_{ab}=\partial_a A_b -\partial_b A_a $, with  four-potential 
\be\label{PotentialVector}
A_{a}=\left(-\frac{Q}{r},0,0,0\right) \,.
\ee
The energy-momentum tensor $T^{(em)}_{ab}$ of the electromagnetic field 
that appears on the right-hand side of (\ref{modifiedEFE}) is then the 
same as the one sourcing the RN metric. However, as for 
the uncharged  Thakurta metric (\ref{NRT-Schwarzschild}), an 
imperfect 
fluid source of the form (\ref{EMTensor}) is now needed in addition to 
the electromagnetic energy-momentum tensor, with a four-velocity $u_a$  
and an energy flux density $q_a$ given by
\begin{eqnarray}
u_{a} &=&  \Big( - a\sqrt{f(r)},0,0,0 \Big) \,, \label{uAndq1}\\
&&\nonumber\\
q_a &=&  \Big( 0, -\frac{\dot{a}(mr-Q^2)}{4\pi 
a^2r^3f(r)^{3/2}}, 0 , 0 \Big) \,. \label{uAndq2}
\end{eqnarray}
The energy density and pressure of such a fluid are 
\begin{eqnarray}
\rho (t,r) &=& \frac{3\dot{a}^2}{8\pi a^4f(r)} \,,\\
&&\nonumber\\ 
P (t,r) &=& \frac{\dot{a}^2-2a\ddot{a}}{8\pi a^4f(r)} \,.
\end{eqnarray}
The energy-momentum tensors of the electromagnetic field and of the 
imperfect fluid appearing on the right-hand side of the Einstein equations 
and responsible for sourcing the metric thus read
\begin{eqnarray}\label{EMImperfectTensors}
T^{(fluid)}_{00} &=& \frac{3\dot{a}^2}{8\pi a^2} \,,\qquad\qquad 
T^{(em)}_{00} =\;\frac{Q^2f(r)}{8\pi r^4} \, ,\nonumber\\
&&\nonumber\\
T^{(fluid)}_{11}& =& \frac{\dot{a}^2-2a\ddot{a}}{8\pi a^2f(r)^2} \,,\qquad 
 T^{(em)}_{11} = -\frac{Q^2}{8\pi r^4f(r)} \,,\nonumber\\
&&\nonumber\\
T^{(fluid)}_{22}& = &\frac{r^2(\dot{a}^2-2a\ddot{a})}{8\pi a^2f(r)} 
\,,\qquad 
 T^{(em)}_{22}= \frac{Q^2}{8\pi r^2} \,,\nonumber\\
&&\nonumber\\
T^{(fluid)}_{33}& = & T^{(fluid)}_{22}\sin^2\theta \, ,\qquad 
 T^{(em)}_{33} = \; T^{(em)}_{22}\sin^2\theta \,,\nonumber\\
&&\nonumber\\
T^{(fluid)}_{01}&=& \frac{\dot{a}(mr-Q^2)}{4\pi ar^3f(r)} \,.
\end{eqnarray}
The origin of this imperfect fluid is, again,  the 
non-vanishing second derivative $\tilde{\nabla}_{0} 
\tilde{\nabla}_{1}a(t)$ of the time-dependent conformal factor $a(t)$.  
Also, as for the non-charged Thakurta 
metric~(\ref{NRT-Schwarzschild}), a 
spacetime singularity arises at $r=m\pm\sqrt{m^2-Q^2}$,  given that the 
Ricci scalar is
\be\label{ChargeThakurtaRicci}
\mathcal{R} = \frac{6 \left( \dot{H}+H^2 \right)}{a^2f(r)}\,.
\ee
Since the charged Thakurta metric (\ref{ChargedThakurta}) is 
conformal to  the  Reissner-Nordstr\"om metric, which solves the 
vacuum Maxwell equations  $\nabla_a F^{ab}=0$ and 
$\nabla_{[a} F_{bc]}=0$, one is also 
guaranteed to satisfy the Maxwell 
equations. This can be understood from our discussion 
of Sec.~\ref{sec:2} about the conformal invariance of such equations. 
Physically, although there 
is an induced effective matter creation and flow, the latter is uncharged, as  
shown by the expressions (\ref{uAndq1}), (\ref{uAndq2}). Therefore, the 
Maxwell equations are preserved due to the absence of induced currents.

\subsection{Cauchy horizon and charged Thakurta geometry}
\label{sec:CauchyHorizonNRT}

As is well known \cite{Poisson}, the RN geometry has a Cauchy horizon nested 
inside an event horizon, with radii
\be
r_{\pm}= m\pm \sqrt{m^2-Q^2} \,;
\ee
these horizons are null surfaces \cite{Poisson}. Since the null  
structure is left unchanged  by conformal transformations, one would 
expect these two  null horizons to be mapped into null horizons of the 
CNRT geometry (\ref{ChargedThakurta}), but  they are mapped into null 
spacetime singularities instead. In fact, the  Ricci scalar of the metric 
(\ref{ChargedThakurta}) is given by~(\ref{ChargeThakurtaRicci})  and it 
diverges\footnote{This fact was noted in \cite{RodriguesZanchin}.} 
at the would-be Cauchy and event horizons $r=r_{\pm}$. The main point here 
 is that the Cauchy horizon of the RN black hole  
disappears by embedding it into a non-static FLRW universe.  The conformal 
transformation from the RN to the CNRT black hole brings an 
improvement if $|Q| \leq m$. It is well known \cite{Poisson} that, at 
small radii, 
the RN geometry 
exhibits a negative energy, as measured by the Misner-Sharp-Hernandez 
mass.  In spherical symmetry, the Misner-Sharp-Hernandez mass 
$M_{_{MSH}}$ contained in a sphere of areal radius $R$ is \cite{MSH1,MSH2}
\be
1-\frac{2M_{_{MSH}}}{R}=\nabla^c R \nabla_c R \,.
\ee
For the RN black hole, this quantity is
\be
M_{_{MSH}}= m -\frac{Q^2}{2r} 
\ee
and it is negative for small radii $r< Q^2/(2m)$. The 
Misner-Sharp-Hernandez mass 
of the CNRT geometry is computed either directly or by using the 
transformation property under conformal transformations \cite{AngusEnzo1,AngusEnzo2} 
\be
\tilde{M}_{_{MSH}}= \Omega M_{_{MSH}}-\frac{R^3}{2\Omega} \, \nabla^c\Omega 
\nabla_c \Omega -R^2 \nabla^c \Omega \nabla_c R \,.
\ee
In either way, one obtains for  CNRT
\be
\tilde{M}_{_{MSH}}= a\left( M_{_{MSH}}+  \frac{ 3aH^2 r^3}{2f} 
\right) \,.
\ee
This quantity is non-negative in the entire physical range $r> r_{+}$ 
if $|Q|\leq m$. In fact,  since for the RN black hole $M_{_{MSH}}\geq 0$ 
when $r \geq m/2$, it 
follows that $\tilde{M}_{_{MSH}} >0 $ for any $r> r_{+}$. In the 
supercritical case $|Q|>m$ in which the RN geometry contains a naked 
singularity, instead, $\tilde{M}_{_{MSH}}$ becomes arbitrarily negative at 
small radii (the physical range of values of the radial 
coordinate is now $r>0$).

\subsection{Apparent horizons}
\label{subsec:C}

The areal radius of the CNRT geometry~(\ref{ChargedThakurta}) is 
\be
R(t, r)=a(t) r \,,
\ee
and, as usual in spherical symmetry, the AH radii are 
located  by the roots of the 
equation \cite{MSH1,MSH2,mylastbook} 
\be
\nabla^c R \nabla_c R=0 \,. \label{eq:AH}
\ee
Since
$\nabla_c R = \dot{a} r \delta_{c 0} + a \delta_{c 1} $, this 
equation corresponds to 
\be
\nabla^c R \nabla_c R  =    
\frac{1}{f} \left( f^2 -H^2 r^2 \right)   =0
\,, \label{100-9}
\ee
where $H \equiv \dot{a}/a$ is the 
Hubble parameter.  Since $r>r_{+}$,  we 
have $f>0$ and, taking 
the positive sign in the square root of 
Eq.~(\ref{100-9}), the AHs correspond to the roots of 
\be
f(r) \equiv 1-\frac{2m}{r}+\frac{Q^2}{r^2} = Hr > 0 \,, 
\label{200-10}
\ee
therefore it is clear that the AHs (when they exist) do not 
coincide with the null spacetime singularities at $r_{\pm}$ (which 
correspond to $f=0$ instead). Equation~(\ref{200-10}) corresponds to the 
cubic 
\be
H r^3 -r^2 +2mr -Q^2 =0 \,, \label{200-11}
\ee
but it is more interesting to discuss Eq.~(\ref{200-10}) graphically. The 
AHs correspond to the intersections between the graph of the function  
$y=f(r)$ and the straight line $y=H r$. A qualitative graphical  
analysis determines when roots exist and the number of these roots lying 
in the physical region $r>r_{+}$.  Since 
\be
f'(r)= \frac{2}{r^2} \left( m-\frac{Q^2}{r} \right) \,,
\ee
the function  $f(r)$, which tends to $+\infty$ as $r\rightarrow 0^{+}$,  
decreases for $0<r< r_{min}$, has a minimum 
$f_{min}=1-m^2/Q^2$  at $r_{min} =Q^2/m$, and it increases for $r> 
r_{min}$, asymptoting to~1 as $r\rightarrow +\infty$. 
For reference, we consider 
in all cases a FLRW universe which begins with a Big Bang at which the 
Hubble parameter $H$ diverges and expands for an infinite time. For 
definiteness, we use a dust-dominated FLRW universe with scale factor 
$\left( t/t_0  \right)^{2/3}$.  Then, the  slope $H(t)$ of the 
straight line through the origin $y=Hr$ decreases  as 
time evolves,  from positive infinity near the Big Bang to zero as 
$t\rightarrow +\infty$. We discuss 
separately the 
subcritical, critical, and supercritical situations $|Q|<m$, 
$|Q|=m$, and $|Q| > m$, respectively.

\subsubsection{$|Q|< m$}

\begin{figure} 
\includegraphics[width=6.5cm]{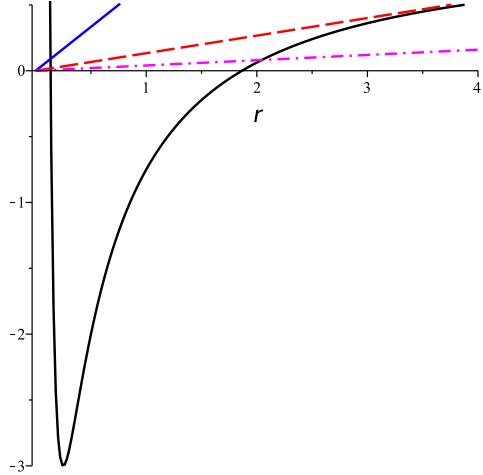}
\caption{The intersections between the curves $y=f(r)$ and $y=Hr$ 
corresponding to the AHs for $m>|Q|$. The solid, dashed, 
and dash-dotted straight 
lines correspond to progressively larger and larger times.  
\label{fig:intersections1}}  
\end{figure}

When the CNRT metric is conformal to a subcritical RN black 
hole,  
the function $f(r)$ vanishes at $r_{\pm}=m \pm \sqrt{ m^2-Q^2}$ and 
its minimum $f_{min}= 1-m^2/Q^2$ is negative. There are three 
possibilities, reported in Fig.~\ref{fig:intersections1} (which is drawn 
for the parameter choice $|Q|=m/2$, $t_0=5m$). 

The straight line  $y=Hr$ intersects the curve $y=f(r)$ at only one 
point if the slope $H(t) $ is sufficiently large, that is, at early 
times near the Big Bang. This 
intersection 
corresponds to the unphysical region $r<r_{-}$ and there are no AHs.

\begin{figure} 
\includegraphics[width=6.5cm]{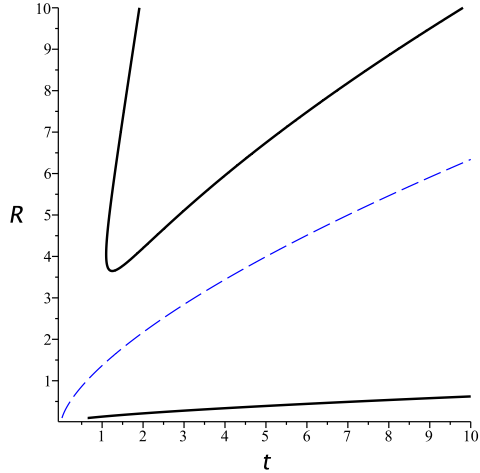}
\caption{The AHs areal radii as functions of the FLRW 
comoving time for 
$|Q|<m$ (``C-curve'' phenomenology). The dashed line describes the 
null singularity at $R_{+}$ and 
the third AH below it is irrelevant for the 
spacetime corresponding to $r>r_{+}$.  \label{fig:AH1}}  
\end{figure}

As time goes by, the slope of the straight line $y=Hr$ decreases 
and the latter
 becomes tangent to the curve $y=f(r)$ at a critical time 
$t_*$, at which a pair of AHs is created.  These AHs necessarily have a 
radius $r_*> r_{+}$, as is clear from 
Fig.~\ref{fig:intersections1}. This critical situation occurs when the 
slopes of the 
straight line and of the curve $f(r)$ coincide, $f'(r)=H(t)$, or 
\be
H r^3 -2m r +2Q^2=0 \,.
\ee
As time progresses ($t>t_{*}$) the two roots separate, 
becoming two distinct intersections between the two 
curves, which correspond to two distinct AHs of radii 
$r_{1,2}$ (labelled so that  $r_2>r_1$). As time grows and $t\rightarrow 
+\infty$, the line $y=Hr$ becomes closer and closer to the 
horizontal and the 
smallest root $r_1 \rightarrow r_{+}$, while $r_2 \rightarrow +\infty$. 
The AH corresponding to the largest root $r_2$ is 
cosmological and $r_2$  reduces to the radius of the cosmological AH of the 
spatially flat FLRW  universe $r_2 \approx 1/H $ (or 
$ R_2\approx a/H $) as $ r_2 \rightarrow +\infty$, which happens as 
$t\rightarrow + \infty$. The smaller root $r_1$ corresponds to a black 
hole AH  which always covers the null spacetime singularity 
($r> r_{+}$) but approaches it as $t \rightarrow +\infty$. The behaviour 
of 
the areal radii of these AHs  versus the comoving time of the 
``background'' universe is given 
in Fig.~\ref{fig:AH1} for  the parameters choice $|Q|=m/2$, 
$t_0=5m$. This phenomenology of AHs is well known 
and is dubbed ``C-curve'' 
in the literature \cite{mylastbook}.

\subsubsection{$|Q|= m$}

\begin{figure} 
\includegraphics[width=6.5cm]{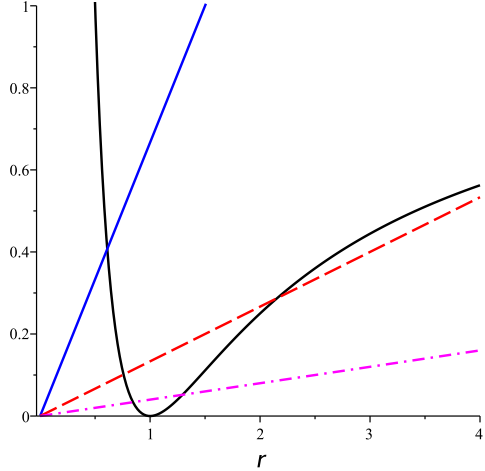}
\caption{The intersections between $y=f(r)$ and $y=Hr$ for 
$|Q|=m$. There are no AHs at early times (solid 
straight line), 
then a pair of AHs appears (dashed line). The cosmological 
one expands forever, while the black hole one shrinks and approaches the 
singularity 
at $R_{+}$ as $t\rightarrow +\infty$ (dash-dotted line). 
\label{fig:intersections2}}  
\end{figure}

In this case the CNRT metric is conformal to an extremal RN black hole in 
which Cauchy and event horizons coincide. The 
null spacetime singularities of the CNRT geometry at $r_{\pm}=m$ 
coincide and there are only two spacetimes disconnected by it. Now the 
function 
\be
f(r)=\left( 1-\frac{m}{r} \right)^2 
\ee
is non-negative and vanishes only at its minimum, achieved at 
$r_{\pm}=m$. Repeating the graphical analysis (see 
Fig.~\ref{fig:intersections2}), 
at 
early times and high values of $H$, there is only one root $r_1$ 
with $0<r_1 < m $, which lies in the unphysical region, and there are 
no 
AHs. As time reaches a critical value $t_*$, two AHs appear, corresponding 
to a double root $ r_* > m $ and to the 
straight line $y=Hr$ being tangent to the curve $y=f(r)$. At later 
times $t> t_*$, there are two distinct roots $r_{1,2}$ with $m< r_1 < 
r_2$. As 
the universe evolves and $t\rightarrow + \infty$, $r_1 \rightarrow m$ 
and $r_2 
\rightarrow +\infty$. The AH at $r_1$ is interpreted as a 
black hole AH, while the one at radius $r_2$ is interpreted 
as 
a cosmological AH, which approaches the usual FLRW AH of areal radius 
$R_2=a/H $ at late times. Qualitatively, the 
situation is similar to that of the previous case $|Q|< m$.

\subsubsection{$|Q|> m$}

In this case the CNRT geometry is conformal to a RN supercritical solution 
which does not have horizons and exhibits a naked singularity at $r=0$. 
The  physical range of the coordinate $r$ is now the entire interval $ 
r>0$. The function $f(r)$ can be written as  
\be
f(r) =\frac{1}{r^2} \left[ \left( r-m\right)^2 +Q^2 -m^2 \right] 
\ee
and is always positive, with positive minimum 
\be
f_{min}=f\left(  \frac{Q^2}{m} \right) =1-\frac{m^2}{Q^2} \,.
\ee
The equation locating the AHs, $ f(r)=Hr >0$ can still be 
satisfied. Now the 
situation is qualitatively different from the previous cases.

\begin{figure} 
\includegraphics[width=6.5cm]{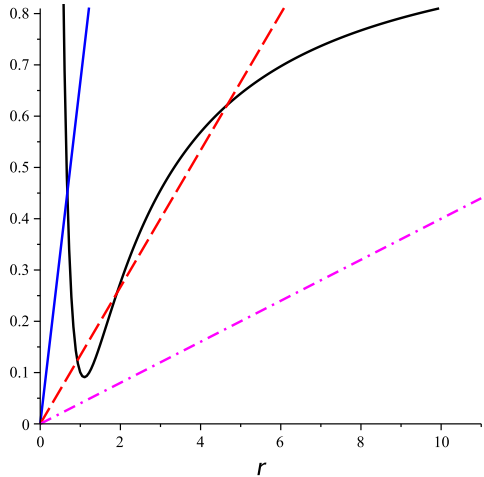}
\caption{The intersections between $y=f(r)$ and $y=Hr$ for 
$|Q|>m$. At early times (solid line) there is only one AH. As time 
progresses (dashed line), a pair of AHs 
appears and  there are three of them. At later times a pair of AHs merge 
and disappear, leaving only a cosmological AH (dash-dotted line). 
\label{fig:intersections3}}  
\end{figure}

\begin{figure} 
\includegraphics[width=6.5cm]{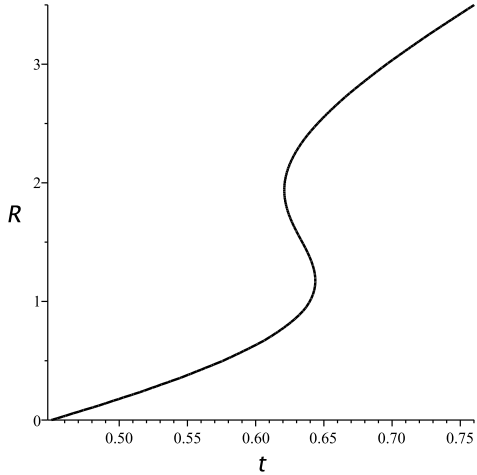}
\caption{The ``S-curve'' phenomenology of the AHs for 
$|Q|>m$. Initially there is only one AH, then a pair of AHs appears, 
one expanding and one shrinking. Later on, two AHs merge and disappear, 
leaving behind only a cosmological AH.  \label{fig:AH2}}  
\end{figure}

Referring  to Fig.~\ref{fig:intersections3} for illustration, one sees 
that at early 
times, when 
the slope of the line $y=Hr$ is large, there is only one root (a 
cosmological AH) in 
the region $r<Q^2/m$: 
this spacetime region hosts a naked singularity. Later on, at a critical 
time $t_1$, we have a single root $r_1<Q^2/m$ and a double root  
$ r_2 > Q^2/m$.  As time progresses, this double root splits in two 
and there are three distinct AHs with radii $r_{1,2,3}$ satisfying  
\be
0< r_1 < \frac{Q^2}{m} < r_2 < r_3 \,.
\ee
As time progresses and the slope of the straight line decreases, 
$r_1$ increases and approaches $ Q^2/m$, while     
$r_2$ decreases approaching $ Q^2/m$, and $r_3$ increases without 
limit. At a  
critical time $t_2>t_1 $, $r_1$ and $r_2$ merge, corresponding to the 
annihilation of these two AHs,  while $r_3$ (corresponding 
to a cosmological AH) still exists. At times $t> t_2$, there 
is only one intersection  between $y=f(r)$ and $y=Hr$, with radius 
$r_3 \rightarrow +\infty$ as $t\rightarrow + \infty$. This surviving 
AH is a nearly-FLRW cosmological AH. At times 
$t> t_2$, the spacetime hosts a naked singularity not covered by a black 
hole AH. The behaviour of the areal radii of the AHs  versus the comoving 
time of the ``background'' universe is given in 
Fig.~\ref{fig:AH2} for the parameter choice $|Q|=3m/2$ and 
$t_0=5m$. 

This behaviour of the AHs is the alternative to the 
``C-curve'' phenomenology most often seen in the literature on AHs, and is 
called ``S-curve'' behaviour \cite{mylastbook}. It was 
found for the first time in the Husain-Martinez-Nu\~nez solution of GR 
sourced 
by a free, canonical and minimally coupled  scalar field \cite{HMN}. The 
appearance of a naked singularity in the supercritical CNRT geometry  is 
not too surprising, since the latter  is conformal to a naked singularity 
spacetime. (This is also the case for the Husain-Martinez-Nu\~nez 
spacetime, which is conformal to the Fisher scalar field solution hosting  
a naked singularity \cite{HMN}.)

\section{Cauchy horizon and McVittie metric} 
\label{sec:5}
\setcounter{equation}{0}

As discussed above, the central assumption made in 
Refs.~\cite{Cardosoetal,Reall} is a charged black hole embedded in a 
static de Sitter background, 
as well as a constant charge assigned to the black hole. 
While the weakness of the latter assumption will be dealt with elsewhere, 
our goal in this section is to deal with the former assumption using 
yet another spacetime representing a black hole embedded in a 
cosmological background. 
Indeed, we know from the Friedmann equation corresponding to an FLRW 
universe that, whenever there is matter, the universe cannot 
describe a de 
Sitter background as the Hubble parameter governed by such an equation can 
never be constant. Then, this fact allows one to argue 
that by embedding the RN black hole in a more ``realistic'' background, 
the Cauchy horizon would always be hidden behind a singularity. It 
 turns out that this is what happens whenever the background 
is allowed to be non-static as is the case for the McVittie spacetime.

The McVittie spacetime (\ref{McVlineelement}) was introduced long ago in order to study the 
competition between cosmic expansion and local dynamics \cite{McVittie}. 
It is regarded as describing a black hole embedded in a FLRW universe and, 
recently,   
it has been the subject of considerable  attention  
\cite{genMcVittieHorndeski,AudreyPRD,McVittierecent1,McVittierecent2,McVittierecent3,McVittierecent4,McVittierecent5,McVittierecent6,McVittierecent7,McVittierecent8,McVittierecent9,McVittierecent10,McVittierecent11,McVittierecent12,McClureDyer2006,RodriguesZanchin}. A charged 
version of 
the McVittie cosmological 
black hole was introduced in \cite{ShahVaidya68}, generalized in 
\cite{MashhoonPartovi79}, and further studied in 
\cite{McClureDyer2006,furtherMcV1,furtherMcV2,furtherMcV3,VFAndresAngus,RodriguesZanchin}. The 
line element and Maxwell field assume the form
\begin{eqnarray}
ds^2 & = & -\frac{ \left[  1-\frac{ \left( m^2-Q^2 \right)}{4a^2r^2} 
\right]^2 }{ \left[ \left( 1+\frac{m}{2ar}\right)^2-\frac{Q^2}{4a^2r^2} 
\right]^2} \,  dt^2+a^2(t)\left[ \left( 1+\frac{m}{2ar} \right)^2 
-\frac{Q^2}{4a^2r^2}\right]^2 \left( dr^2 
+r^2 d\Omega_{(2)}^2 \right)  \label{QMcVlineelement} \\
F^{01} &=& \frac{Q}{ a^3 r^2\left[ 1-\frac{  \left( 
m^2-Q^2\right)}{4a^2r^2} \right]
\left[ \left( 1+\frac{ m}{2ar}\right)^2 -\frac{Q^2}{4a^2r^2}\right]^2} 
\,,
\end{eqnarray}
where the parameters $m>0$ 
and $Q$ describe the mass
and the electric charge, respectively, while 
$a(t) $ is the scale factor of the ``background'' FLRW universe.
The line element (\ref{QMcVlineelement}) interpolates between the 
RN spacetime (obtained for $a \equiv 1$) and the 
spatially flat FLRW metric (obtained for large values of $r$).  The 
geometry reduces to the spatially flat FLRW one if $m=Q=0$.

The AHs of the charged McVittie metric have been studied in 
\cite{VFAndresAngus,RodriguesZanchin}. The areal radius is 
\be
R(t,r) 
= m+a(t)r+\frac{m^2-Q^2}{4a(t)r}, 
\ee 
with $R \geq m$ if $|Q| \leq m$.   The 
Ricci scalar 
\be 
{\cal R} = 6 \left[ 2H^2 +  \dot{H} \left(  \frac{ 1+\frac{m}{ar}+
\frac{\left(m^2-Q^2 \right)}{4a^2r^2}   }{
1- \frac{ ( m^2-Q^2)}{4 a^2r^2}  } \right) \right]   \label{Ricciscalar}
\ee
(where $H (t)\equiv \dot{a}/a$) is singular at 
$R_{*}=m+\sqrt{m^2-Q^2}$ 
if 
$|Q|\leq m$. This is the location of the outer apparent horizon of the RN 
geometry. This spacelike singularity splits the 
spacetime into two 
completely disconnected portions.

The line element of the charged McVittie spacetime in $\left( 
t,R,\theta,\phi \right)$ 
coordinates is
\begin{eqnarray}
ds^2 &=& -\left(1-\frac{2m}{R}+\frac{Q^2}{R^2}-H^2R^2\right)dt^2-\frac{2HRdRdt}{\sqrt{1-\frac{2m}{R}+\frac{Q^2}{R^2}}}+\frac{dR^2}{ 
1-\frac{2m}{R}+\frac{Q^2}{R^2}}+R^2d\Omega^2 \,.\label{tRCoordinates}
\end{eqnarray} 
Equation~(\ref{eq:AH}) locating 
the AHs becomes the quartic 
\be\label{AHDetection}
{\cal G}(t,R)= H^2R^4 -R^2 +2mR -Q^2=0 \,.
\ee
For large radii one obtains the asymptotic root $R \simeq H^{-1}$, which 
corresponds to the cosmological AH of the FLRW background. If 
$H\rightarrow 0$, there are only the two roots $R_{\pm}= m \pm \sqrt{m^2-Q^2}$, where the smaller one, $R_{-}$, is always located inside the 
spherical singularity and $R_{+}$ lies outside of it. In order to locate 
the AHs numerically, one needs to fix the FLRW background. 
However, in order to have a general result, which would be independent of 
the particular conformal factor $a(t)$, we shall first repeat the 
procedure applied on the charged Thakurta metric in 
subsection~\ref{subsec:C} and investigate the occurrence of a null 
internal horizon that we would identify with the Cauchy horizon. After 
locating these various AHs we shall investigate the possibility of 
identifying one of them -- the internal one -- as a Cauchy horizon.

\begin{figure}
\center\includegraphics[width=8.5cm]{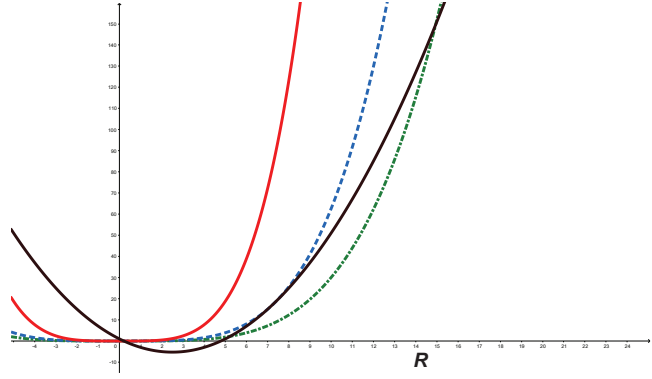}
\caption{The locations of the AHs for 
$|Q|>m$. Initially there is only one AH (intersection of the black 
solid curve 
$y=R^2-2mR+Q^2$ with the red curve $y=H^2R^4$), then a pair of AHs 
appears (intersection of the black solid curve with the blue dashed curve 
$y=H^2R^4$ with larger $H$ at later times), then three AHs emerge 
(intersection of the 
black solid curve 
with the green dash-dotted curve $y=H^2R^4$ at much later times).} 
\label{fig:McVittyAHs}
\end{figure}

The solutions to Eq.~(\ref{AHDetection}) can be found graphically as shown 
in Fig.~\ref{fig:McVittyAHs} by detecting the intersections of the 
parabola 
$H^2R^4$ (shown in red, blue, then green for consecutive times 
corresponding to smaller and smaller values of $H(t)$) with the parabola 
$R^2-2mR+Q^2$ (solid black curve). For different moments in 
the evolution of the universe described 
by the scale factor $a(t)$ we obtain the pattern shown in 
Fig.~{\ref{fig:McVittyAHs}}. 

In order for an AH detected by (\ref{AHDetection}) to be null, its 
normal  $\nabla_a {\cal G}$ needs to satisfy $\nabla_a {\cal 
G} \nabla^a {\cal G} = 0 $. In terms of 
the metric, this reads
\begin{equation}
\label{FormalNullCondition}
g^{RR} \left( \partial_R {\cal G} \right)^2 + 2g^{Rt} \partial_R  {\cal G} 
\, \partial_t 
{\cal G}+g^{tt} \left( \partial_t {\cal G} \right)^2=0.
\end{equation}
Because ${\cal G}(t,R)=0$ at an AH, using (\ref{tRCoordinates}) we 
can compute the needed components of the 
inverse metric: 
\begin{equation}\label{InverseMetric}
g^{RR}=0 \,,\qquad g^{Rt}=- 1 \,,\qquad g^{tt}=- \frac{1}{H^2R^2} \,.
\end{equation}
On the other hand, computing the partial derivatives in (\ref{FormalNullCondition}), we find,
\begin{equation}\label{Partials}
\partial_R {\cal G}=4H^2R^3 - 2R + 2m,\qquad\partial_t {\cal G} 
= 2H \dot{H} R^4 \,.
\end{equation}
Substituting the results (\ref{InverseMetric}) and (\ref{Partials}) in (\ref{FormalNullCondition}), we find the following condition 
for one of the AHs of the McVittie spacetime to be null,
\begin{equation} \label{Eq2}
H\left(4H^2R^3-2R+2 m\right)+\dot{H}R^2=0 \,.
\end{equation}

Thus, we conclude that a  given apparent horizon of the McVittie spacetime 
is not necessarily null. 
Instead, an algebraic 
equation in $R$ and $H$ has to be satisfied. One should keep in mind, 
though, the important fact that, because a Cauchy horizon is a null hypersurface, one needs two equations to be satisfied in order to be able to identify an AH with a Cauchy 
horizon. On one hand, for a given Hubble expansion rate $H$, 
Eq.~(\ref{AHDetection}) gives 
for all 
times $t$ the corresponding location $R$ of the AH. On the other hand, Eq.~(\ref{Eq2}) is required for such a horizon to be null at whatever location it happens to be and at any corresponding time. However, satisfying both equations 
(\ref{AHDetection}) and (\ref{Eq2}) can only happen at a finite number of instants of time $t$ because extracting $t$ in terms of $R$ from the first and then substituting in the second leads to an algebraic equation in $R$ alone and hence can only yield discrete pairs $\left(t_0, R_0 \right)$.

These results show that, whenever the background is not static and not artificially created by a conformal transformation like in a McVittie spacetime, the would-be Cauchy horizon would only exist at certain instants 
of time. In the next subsection, we illustrate the occurrence of the 
various apparent horizons using a concrete model of an expanding universe.

\subsection{AHs in a charged McVittie spacetime with scale factor $a(t)=a_0t^p$}
 
\begin{figure}
\centering
\includegraphics[scale=0.3]{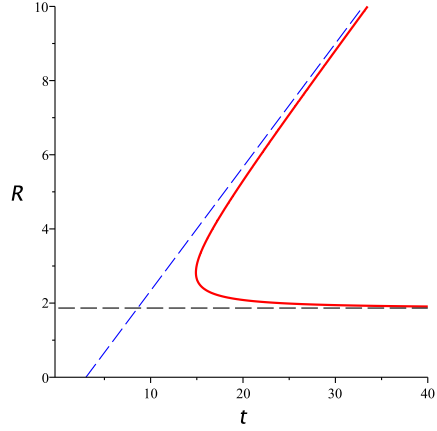}
\caption{The AH radii in the charged McVittie spacetime with FLRW scale factor $a(t)=a_0t^3$. A black hole AH and a cosmological AH are  born at a critical time. The cosmological horizon expands forever, while the black 
hole AH asymptotes to the spacetime  singularity at $R_{*}=m +\sqrt{m^2- Q^2}$ (the horizontal black dashed line). The (blue) dashed, oblique line of equation $R=H^{-1} -m$ is an asymptote  for  the cosmological AH at late times and large radii.}  \label{fig:6}
\end{figure}

\begin{figure}
\centering
\includegraphics[scale=0.3]{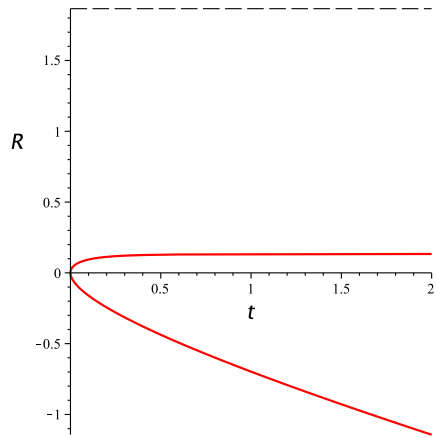}
\caption{The third AH is located in the region below the 
singularity (the dashed horizontal line) and does not belong to the region 
$R>R_*$. A fourth formal 
root of Eq.~(\ref{eq:AH}) is negative and has no physical meaning.}     
\label{fig:7}
\end{figure}

In keeping with the spirit of Ref.~\cite{Cardosoetal}, we choose a 
dark-energy dominated and accelerating FLRW ``background'' with scale 
factor 
$a(t) =a_0 t^p$ with $p>0$ and, as a specific example, $p=3$. The 
behaviour of the AH radii (in units of $m$) are plotted in 
Figs.~\ref{fig:6} and \ref{fig:7}  
as functions of the comoving time $t$ (also measured in units of $m$) for 
the particular 
parameter choice  $Q =\pm m /2$. (For ease of illustration, the  
scale is different in the two figures.) 

Figure~\ref{fig:6} reports the AH radii as given by~(\ref{AHDetection}) 
in the spacetime region $R> 
m +\sqrt{m^2-Q^2}$ above the spacetime singularity. This is again a 
``C-curve'' phenomenology.  There are no 
AHs in this spacetime region at early times. Then, a 
black hole AH and a cosmological AH form as  
a pair at a critical time. The cosmological horizon expands forever, while 
the black hole horizon asymptotes to the spacetime  singularity  at
$R=m+\sqrt{m^2 - Q^2}$ (represented by the black horizontal dashed 
line in  Fig.~\ref{fig:6}). The blue, dashed, oblique line of equation 
$R=H^{-1} -m$ is 
an 
asymptote for  the cosmological AH at late times and
large radii \cite{VFAndresAngus}. Once the two AHs form, there 
is no inner Cauchy horizon for the dynamical cosmological black hole thus 
formed. The singularity coincides with the outer event horizon of the 
RN black hole, which is obtained for $a=1$. The third root of 
Eq.~(\ref{AHDetection}) 
corresponds to an AH located in the other spacetime 
``below'' the singularity 
 $R< m +\sqrt{m^2-Q^2}$.  Figure~2 
reports this third root of Eq.~(\ref{AHDetection}).
 Since the two spacetime regions separated 
by the singularity are disconnected, this third AH has no 
implication, or meaning, for the region above the singularity. As remarked 
in Ref.~\cite{VFAndresAngus}, embedding the RN black hole in a 
time-dependent cosmological 
``background'' (not a locally static de Sitter one) has the 
effect of making the Cauchy horizon disappear. This 
fact is consistent with the known instability of this horizon in the RN 
spacetime \cite{PoissonIsrael90}.

The extremal case $|Q|=m$ can be discussed analytically. In this case 
the singularity is located at $R_*=m$ and Eq.~(\ref{AHDetection}) for 
the AHs can be solved exactly, giving  
\be
R_{AH}^{(\pm)}= \frac{1\pm \sqrt{1-4mH}}{2H} \,.
\ee
In a universe with scale factor $a(t)=a_0t^p$, the cosmological and black 
hole AHs $ R_{AH}^{(+)}$ and 
$ R_{AH}^{(-)}$ are created at the critical time $t_0= 4pm$ and exist for 
all times $t>t_0$. Since $R>m_0$, for $t>t_0$ we have 
\be
m_0 <R_{AH}^{(-)} < R_{AH}^{(+)} < \frac{1}{H} \,.
\ee
Again, no inner black hole Cauchy horizon exists, speaking in  favor of  
restoring determinism to GR.

\section{Conclusions}
\label{sec:6}
\setcounter{equation}{0}

We have investigated cosmological black holes obtained by conformally 
transforming either the neutral Schwarzschild black hole or the charged 
Reissner-Nordstr\"om black hole. The first case consists of the uncharged 
non-rotating Thakurta spacetime, while the second one consists of
 the charged version of this geometry.
The general pattern emerging from 
obtaining 
conformal solutions by using ``seed'' solutions of the Einstein 
equations has 
been studied. The analysis shows that, while the resulting action after 
such a transformation is no longer an Einstein-Hilbert type action but a 
Brans-Dicke action with the pathological Brans-Dicke parameter 
$\omega=-3/2$, the resulting Brans-Dicke field equations may nevertheless 
be interpreted as effective Einstein field equations 
with an effective 
imperfect fluid playing the role of an additional source besides  the 
electrovacuum associated with the seed \cite{Pimentel89A,Pimentel89B}. In fact, our whole procedure is based, not on any kind of Brans-Dicke theory, but rather on pure general relativity. The Brans-Dicke theory only suggests itself as the natural framework for interpreting the resulting action and equations of motion. Indeed, our approach, which consists in embedding a black hole in an expanding universe filled by fluids, takes its full meaning in Einstein's gravity. The 
imperfect character of the 
induced effective fluid is manifested by the emergence of an energy 
flow and is unavoidable as long as the chosen embedding 
background 
is evolving, {\textit{i.e.}}, the scale factor is time-dependent. This 
pattern is general and arises for both charged and uncharged black 
holes embedded in cosmological ``backgrounds'' obtained by this conformal 
technique. However, while an 
induced  effective energy flow is automatically obtained even if it is 
absent in 
the original seed solution, no charged flow emerges even 
for  charged black hole seeds, as a consequence of the conformal 
invariance of the vacuum Maxwell equations.

This technique has then been used to tackle the problem of determinism in 
GR by building the charged cosmological 
black hole in the 
form of the charged non-rotating Thakurta spacetime. Such a spacetime 
requires the presence of a neutral, but imperfect, fluid as a source. 
Nevertheless, the corresponding black hole is more realistic than the 
RNdS black hole as the former might be chosen to be embedded in a FLRW 
universe, in 
contrast to the latter which lives in a de Sitter ``background''. We found 
that the Cauchy horizon of such a spacetime always hides, in the 
non-extremal case, behind a singularity.

The same analysis has been performed on another type of charged black 
hole embedded in a cosmological ``background'', the 
McVittie geometry. 
We found that whenever the background is not static  nor artificially 
created by a conformal transformation as in the case of the charged 
non-rotating Thakurta spacetime, the would-be Cauchy horizon appears 
only at certain instants of time. Hence, those locations 
could not really qualify as the loci of a real Cauchy horizon that would 
put determinism within GR in jeopardy. Indeed, Cauchy horizons 
are necessarily null hypersurfaces \cite{Waldbook}, whereas the 
would-be inner Cauchy horizon of McVittie spacetime is mainly a non-null 
AH, except at discrete instants of time. 

Now, one might wonder whether the various theoretical models used here could be of any practical 
use in real cosmology, or are they just tools for investigating fundamental theoretical issues, 
such as the loss of determinism in GR. All three of the models presented here, the non-charged 
and the charged non-rotating Thakurta spacetimes, as well as the McVittie spacetime, are pure 
theoretical constructs intended for creating a physical setting that would be as close to the 
real systems of nature as possible. They satisfactorily take into account the fact that real 
black holes in nature represent inhomogeneities, with induced backreactions, in an otherwise 
smooth and homogeneous expanding universe.  We cannot claim, however, that our models could have 
real astrophysical objects as counterparts in nature. The first two models, although based on 
the more ``realistic" Schwarzschild black hole, cannot be used in real cosmology because they 
emerge from a pure theoretical technique. In fact, while the end product of such a technique is 
closer to the real black holes in our expanding universe, the byproduct that comes with such a 
construct is, as we saw, the presence of an artificial imperfect fluid. The third model is also 
a pure theoretical black hole which is, by construction, already embedded in an expanding 
universe. In contrast to the first two models, however, the third one is conceptually more 
attractive as it is free of any artificial fluids. Yet, the very absence of any kind of fluid 
around such a black hole prevents, in turn, the model from totally representing a realistic 
system of nature. Therefore, from this perspective the first two models become, precisely thanks 
to the presence of their imperfect fluids (although artificial) more ``realistic". For 
inhomogeneous exact solutions with perfect fluids, let alone those that lack fluids in their 
background, are of very limited use in astrophysics. In fact, heat fluxes and/or anisotropic 
stresses, supplied only by imperfect fluids, are needed in any kind of simulation and/or 
representation of realistic astrophysical objects in nature. The full merit of all three models 
is thus to investigate and test the formalism of GR in settings that are closer to the real 
world. This, indeed, helps greatly avoiding eventual traps coming from general conclusions --- 
here 
concerning determinism in GR --- based on models that are far from resembling our real expanding 
universe.

Now, the use of the charged McVittie spacetime (excluding the case in which it 
reduces to the RNdS space for $H=$~const.) has of course conceptual weaknesses too. 
First, before the critical time at which the black hole/cosmological 
AH  pair is created, there is a naked 
singularity and this 
solution of the Einstein equations cannot be obtained as the development 
of regular Cauchy data. Second, AHs ultimately depend on the 
foliation \cite{WaldIyer1,WaldIyer2}, although all spherically symmetric foliations 
(the only ones of practical importance here) determine the same AHs 
\cite{Ellis}. Third, while the RN solution is the most general 
spherical and locally static electrovacuum solution of the Einstein 
equations with positive $\Lambda$, in the presence of a fluid there is no 
general solution and the charged Thahurta and McVittie geometries cannot 
claim such a 
degree of generality. In spite of these shortcomings, a time-dependent 
FLRW ``background'' is a more general setup for a charged black hole than 
the de Sitter one. Once (local) staticity is removed, there is no trace of 
inner Cauchy horizons in charged black holes, according to our models. 
This result agrees with the results of Refs.~\cite{Hod1,Hod2,Hod3,Diasetal} which restore 
determinism to Einstein theory. Given the importance of the issue, 
however, one still hopes that an  experimental test, analogous to the one 
proposed in Ref.~\cite{Virbhadra} to test the Weak Cosmic Censorship 
Hypothesis, could also be devised to test the Strong Cosmic Censorship 
Hypothesis and bring a decisive conclusion about the fate of determinism 
in GR.

\section*{Acknowledgments}

The authors are grateful to the anonymous referee for the pertinent remarks that helped improve the clarity of the manuscript. D.K.C. thanks the Scientific and Technological Research Council of Turkey 
(T\"{U}B\.{I}TAK) for a postdoctoral fellowship through the Programme 
B\.{I}DEB-2219 and Nam{\i}k Kemal University for support. F.H. and V.F. 
are supported by the Natural Sciences and Engineering Research Council of 
Canada (Grants No.~2017-05388 and No.~2016-03803), and all authors thank Bishop's University.

\end{document}